\documentstyle[prl,aps,twocolumn]{revtex}

\newcommand{\be}{\begin{equation}}
\newcommand{\ee}{\end{equation}}
\newcommand{\bea}{\begin{eqnarray}}
\newcommand{\eea}{\end{eqnarray}}
\newcommand{\bref}[1]{(\ref{#1})}
\newcommand{\nn}{\nonumber}
\newcommand{\A}{\alpha} \newcommand{\B}{\beta} 
 \newcommand{\D}{\delta} 
\newcommand{\ep}{\epsilon} 
\newcommand{\T}{\theta} 
   \newcommand{\vp}{\varphi}
         \newcommand{\lam}{\lambda}
           \newcommand{\s}{\sigma}
          \newcommand{\w}{\omega}
          
\newcommand{\h}{\eta}           
           
\newcommand{\W}{\Omega}

\def\6{\partial}
\def\7{\tilde}
\def\8{\hat}

\def\pa{\partial}

\def\CL{{\cal L}}
\def\CE{{\cal E}}

\def\l{{\ell}}
\def\lag{Lagrangian }\def\lags{Lagrangians }
\def\ham{Hamiltonian }

\begin{document}
\title{
{\normalsize\begin{flushright}
CERN-TH/2000-186, LFM(SCF)-00-1,
UB-ECM-PF-00/08,TOHO-FP-0067\\
hep-th/0006235
\end{flushright}}
Hamiltonian Formalism for Space-time Non-commutative Theories
}
\author{
Joaquim Gomis\,$^{1,2}$, Kiyoshi Kamimura\,$^{3}$, Josep Llosa\,$^4$}

\address{
$^1$ Theory Division, CH-1211 Geneva 23, Switzerland\\
$^2$ Departament ECM, Facultat de F\'{\i}sica, Universitat de Barcelona and 
Institut de F\'{\i}sica d'Altes Energies,\\
Diagonal 647,E-08028 Barcelona, Spain\\
$^3$ Department of Physics,
Toho University, Funabashi, 274-8510 Japan\\
$^4$ Departament de F\'{\i}sica Fonamental, Universitat de Barcelona,
Diagonal 647,  
E-08028 Barcelona, Spain
\\[1.5ex]
\begin{minipage}{14cm}\rm\quad
Space-time non-commutative theories are non-local in time. 
We develop the \ham formalism for  non-local field theories in $d$ space-time 
dimensions by considering auxiliary $d+1$ dimensional field theories 
which are local  with respect to the 
evolution time. The Hamiltonian path
integral quantization is considered and 
the Feynman rules in the  Lagrangian formalism are derived. 
The case of non-commutative $\phi^3$ theory is considered as an example.
\\[1ex]
PACS numbers: 11.10.Ef, 11.10.Lm, 11.15.Kc
\end{minipage}
}

\maketitle

\section{Introduction}

Space-time non-commutative field theories have  
peculiar properties due to their acausal behavior 
\onlinecite{sst,ab} and lack of unitarity \cite{gm}.
In reference\cite{agm} 
it has been shown that there is a  relation 
between lack of unitarity and the 
obstruction to finding a decoupling limit of string theory in an
electromagnetic background
\onlinecite{sw,sst1,gms,gmms,br,km}. 
These theories have an infinite number of temporal and spatial derivatives,
and therefore are non-local in time and space\onlinecite{fn1,fn3}.
The initial value problem of a non-local theory 
requires to give a trajectory or a finite piece of it\cite{fn2}. 
The Euler-Lagrange  (EL) equation is a constraint 
in the space of trajectories.

The Hamiltonian formalism for non-local theories was presented in \cite{lv}. 
In this paper we improve
the formalism by clarifying the relation among
the Lagrangian and Hamiltonian structures.
We first consider an equivalent theory 
in a space-time of one dimension higher than that of the  original
theory. This space has 
"two times" 
and the dynamics is described in such
a way that the evolution is local with respect to one of the times.
For this equivalent theory one can construct the Hamiltonian. 
A characteristic feature of the \ham formalism for non-local theories
is that it contains the EL equations as \ham constraints.

The Hamiltonian path integral for the $d+1$ dimensional field theory
is constructed. 
The Lagrangian path integral 
formalism for the $d$ dimensional theory is obtained by
integrating out the momenta.

We apply  the \ham formalism to
time-like  and light-like
non-commutative theories\cite{agm}. 
As an example we consider the case of non-commutative $\phi^3$ theory
in $d$ dimensions with space-time non-commutativity.
The action contains the free Klein-Gordon
\lag and the interaction \lag 
$L_i=-\frac {g}{3!} \int d\vec x~\phi*\phi*\phi$\,,
where $*$ refers to the Moyal product. 
We construct the \ham  in $d+1$ dimensions. 
In the path integral  quantization
we get the Feynman rules that 
coincide with those used in references \onlinecite{sst,ab,gm}. 
The theory is unitary at the classical level 
(tree level) but is not unitary at one loop \cite{gm}.
This analysis should shed new light on the structure of these theories. 
The knowledge of the \ham of time-like and light-like non-commutative
field  theories could also be useful to study the energy of their solitons.

\section{ Euler-Lagrange equations for non-local theories}

Unlike standard Lagrangians, that depend on the values of a finite number
of derivatives at a given time ---$q(t)$, $\dot q(t)$, \ldots $q^{(n)}(t)$---,
a non-local \lag depends on a whole piece of the trajectory
$q(t+\lambda)$, for all values of $\lambda$, that is, 
$ L^{\rm non}(t) = L([q(t+\lam)])$. 
At best it can be written as a
function of all time derivatives $q^{(j)}(t)\,$, $j=0,1,2,\ldots$ at the
same $t$. 
This means that the analog of the tangent bundle for 
\lags depending on positions and velocities is infinite dimensional.
The action is
\be
S[q]=\int dt~ L^{\rm non}(t). 
\label{action}
\ee

The EL equation is obtained as the variation of functional \bref{action} and is
given by
\be
\int dt~E(t,t';[q])=0,
\label{EL}
\ee
where 
$ \displaystyle{E(t,t';[q])=\frac {\delta L^{\rm non}(t)}{\delta q(t')}}$.

The EL equation  
must be understood as a functional relation to be fulfilled by 
physical trajectories. It is not a differential system, as one is used
to find for local Lagrangians. In the latter case, the theorems of
existence and uniqueness of solutions enable to interpret the EL equation as
ruling the time evolution of the system, whose state at every instant of
time is represented by a point in the space of initial data, {\it e.g.},
$J_L=\{q,\dot q,\ldots, q^{(2n-1)}\}$ for a local Lagrangian of order $n$.

In the non-local case, if we denote 
the space of all possible trajectories as
$J=\{q(\lambda), \lambda\in R\}$, (\ref{EL}) is a Lagrangian constraint 
defining the subspace $J_R\subset J$ of physical trajectories.

\section{ $1+1$  dimensional field theory 
description of non-local theories}
 
Nevertheless, if we insist in defining a ``time evolution" ${T_t}$
for a given initial trajectory $q(\lam)$, a natural choice is:
\be
 q(\lambda) \, \stackrel{T_t}{\longrightarrow} \,q(\lambda+t) \,.
\label{timeEv}
\ee
We shall hence introduce new dynamical variables $Q(t,\lambda)$ such that
\be
Q(t,\lambda) = q(\lambda + t)\,.
\label{timeEv1}
\ee
Thus, $t$ is the ``evolution" parameter and $\lam$ is a continuous parameter 
indexing the degrees of freedom. 
These new variables follow the 
evolution (\ref{timeEv1}) above, and $Q(0,\lambda)$ can be seen as initial 
data in the local $1+1$  dimensional field theory. 

In differential form, condition (\ref{timeEv1}) reads:
\be
\dot Q(t,\lam) = Q^\prime(t,\lam),
\label{Qdot}
\ee
where `dot' and `prime' respectively stand for $\partial_t$ and
$\partial_\lambda$. 

We then consider the Hamiltonian system for the  $1+1$ dimensional  field $Q$ 
with the \ham
\be
H(t,[Q,P])=\int d\lam P(t,\lam)Q^\prime(t,\lam)-\tilde L(t,[Q]),
\label{h}
\ee
where $P$ is the canonical momentum of $Q$. The phase space is thus $T^\ast J$
with the fundamental Poisson brackets 
\be
\{Q(t,\lam),P(t,\lam')\} = \delta(\lam-\lam') \,.
\label{PB}
\ee
In the \ham \bref{h}  $\tilde L(t,[Q])$ is a functional defined by
\be
\tilde L(t,[Q]):=
\int d\lam\,\delta (\lam) {\cal L}(t,\lam)\,.
\label{lagr1}
\ee
The ``density" $\CL(t,\lam)$ is constructed from $L^{\rm non}(t)$ by replacing
$q(t)$ by $Q(t,\lam)$, the $t$-derivatives of $q(t)$ by $\lam$-derivatives 
of $Q(t,\lam)$ and $q(t+\rho)$ by $Q(t,\lambda+\rho)$. 
 In this construction of the \ham $\lam$ inherits the signature 
of the original time $t$ and is a time-like
coordinate. Furthermore the symmetry of the original \lag is realised
canonically in the enlarged space \cite{gkm}.
Note that $\CL(t,\lam)$ is local in $t$ and is non-local in $\lam$.
$H$ depends linearly on $P(t,\lam)$ but 
does not depend on $\dot Q(t,\lam)$.

The relation (\ref{Qdot}) naturally arises as the first Hamilton equation for
(\ref{h}). However there is no a priori relationship between
$P(t,\lam)$ and $Q(t,\lam)$ ---unlike it happens in the local case---, the
second Hamilton equation:
\be 
\dot P(t,\lam) = P^\prime (t,\lam)+\frac{\D \tilde L(t,[Q])}{\D Q(t,\lam)}
\label{Pdot}
\ee
does not imply any further restriction on $Q(t,\lam)$. Thus, the Hamiltonian
system (\ref{h}) on $T^\ast J$ is not so far equivalent to the non-local
Lagrangian system of $L^{non}(t)$.

Now, instead of taking the whole phase space $T^\ast J$, we shall restrict  
to the subspace defined by the 1-parameter set of primary 
constraints \cite{lv}:
\be
\varphi(t,\lam,[Q,P]) \equiv P(t,\lam) - F(t,\lam,[Q]) \approx 0
\label{mom}
\ee
with 
\be
F(t,\lam,[Q]):= \int \,d\sigma \, \chi(\lam,-\sigma) \,\CE(t;\s,\lam) \,,
\label{constr1}
\ee
where $\CE(t;\s,\lam)$ and $ \chi(\lam,-\s)$ are defined by
\bea
\CE(t;\s,\lam)&=&\frac{\D \CL(t,\s)}{\D Q(t,\lam)},~~~
\chi(\lam,-\s)~=~\frac{\ep(\lam)-\ep(\s)}{2}.
\eea
Here $\ep(\lam)$ is the sign distribution. The symbols "$\equiv$" and 
"$\approx$" respectively stand for ``strong" and ``weak" equality.

Further constraints are generated by requiring the stability of
the primary ones. In the first step, we obtain:
$$ \dot\varphi(t,\lam,[Q,P]) \equiv 
\varphi^\prime(t,\lam,[Q,P]) + \delta(\lam)\,\psi_0(t,[Q]) \approx 0 $$
where 
\be
\psi_0(t,[Q]) := \int d\s~\CE(t;\s,0) \approx 0
\label{constr2}
\ee
is the secondary constraint. Further constraints then follow by
successive time differentiations of $\psi_0$. They can be written all together 
in a
condensed form as:
\be
\psi(t,\lam,[Q]) \equiv \int d\s~\CE(t;\s,\lam) \approx 0 \,.
\label{constr3}
\ee
Therefore, the constrained Hamiltonian system defined by the \ham
(\ref{h}) and the primary constraints (\ref{mom}) lives in a reduced phase 
space $\Gamma \subset T^\ast J$ defined by (\ref{mom}) and (\ref{constr3}).
Taking into account (\ref{timeEv1}), the constraint (\ref{constr3}) reduces 
to the EL equation (\ref{EL}) obtained from $L^{\rm non}(t)$. 

The constraints \bref{mom} and \bref{constr3} belong to
the second class in non-singular systems.  In the next section we will show
explicitly, for (non-singular) higher derivative \lag system of order $n$,
they are used to reduce the phase space to  $2n$ dimensions  
reproducing the canonical Ostrogradski formalism \cite{o}.
Our formalism developed here turns out to be a generalization of the
Ostrogradski formalism to the case of infinite order
derivative theories.  
The infinite chain of second class constraints has also appeared in the 
description of boundary conditions as constraints \cite{sjs}.
Summarizing, the equivalence has been built in the $1+1$  dimensional
\ham formalism of {\it local} field theories 
through the constraints \bref{mom} and \bref{constr3}. 
This type of equivalence between the Hamiltonian and Lagrangian formalism is
different from the one in local theories\cite{bgpr}.

\section{ Non-singular Higher Order Derivative theories} 

Here we would like to derive both the Lagrangian and Hamiltonian 
formalisms for non-singular higher order derivative theories from the
Hamiltonian formalism of non-local theories developed in the last 
section\cite{LV}.

Let us consider a regular higher derivative theory described by the
Lagrangian $L(q,\dot q,\ddot q,...,q^{(n)})$ and write the expressions
obtained in the previous section for the  non-local Lagrangian. As we embed
the higher order theory in the non-local setting
we start with the infinite dimensional
phase space 
$T^*J(t)=\left\{ Q(t,\lambda), P(t,\lambda)\right\}$.
They are assumed to be expanded in the Taylor basis \cite{Marnelius} as
\bea
Q(t,\lam) &\equiv&\sum_{m=0}^{\infty}~e_m(\lam)~q^m(t),\;~~~~
\nn\\
P(t,\lam) &\equiv&\sum_{m=0}^{\infty}~e^m(\lam)~p_m(t),
\label{momentahigh}
\eea
where $e^\l(\lam)$ and $e_\l(\lam)$ are orthonormal basis 
\bea
e^\l(\lam)&=&(-\pa_\lam)^\l\D(\lam),~~~~e_\l(\lam)~=~\frac{\lam^\l}{\l!}.
\eea
The coefficients in \bref{momentahigh} are new canonical variables
\bea
\{q^m(t),~p_n(t)\}&=&{\D^m}_n
\label{symplectich}
\eea
and the Hamiltonian \bref{h} is
\bea
H(t)
&=&\sum_{m=0}^{\infty}~p_m(t)~q^{m+1}(t)~-~L(q^0,q^1,...,q^n).
\label{hh}
\eea

The momentum constraint \bref{mom} becomes an infinite set of constraints 
\bea
\vp_m(t)&=&p_m(t)-\sum_{\l=0}^{n-m-1}
(-D_t)^{\l}~\frac{\pa L(t)}{\pa q^{\l+m+1}(t)}\approx 0,
\label{vpmm}
\eea
where
\bea
D_t=\sum_{r=0} q^{r+1}\frac{\pa}{\pa q^r}.
\label{defD}
\eea
On the other hand the constraint $\psi$  in \bref{constr3}
in terms of the Taylor basis becomes 
\bea  
\psi^m(t)&\equiv&(D_t)^m\;[\sum_{\l=0}^n~ (-D_t)^\l
 \frac{\pa L(t)}{\pa q^\l(t)}]\approx 0.
\label{EOMm}
\eea

These constraints \bref{vpmm} and \bref{EOMm} are 
second class and are used to reduce the infinite dimensional phase space to
finite one leading to the ordinary Ostrogradski \ham formalism.
The operator $D_t$ defined in \bref{defD} becomes
a time evolution operator for $q$'s using the first set of Hamilton equation
\bea
\dot q^r&=&q^{r+1}.
\label{qmdot}
\eea
Using this in \bref{vpmm} the constraints 
 $\vp_{m},\;(0\leq m \leq n-1) $ coincide with the definition of
the Ostragradsky momenta
\bea
p_m&\sim& ~\sum_{\l=0}^{n-m-1}
(-\pa_t)^{\l}~\frac{\pa L(t)}{\pa(\pa_t^{\l+m+1} q(t))},
\nn\\&&\hskip 30mm
\;\;(0\leq m \leq n-1).
\label{vpmm2}
\eea
Now  they 
can be solved for $q^\l,\;(n\leq \l \leq 2n-1)$ as functions of
canonical paires $\{q^j,p_j\},\;(0\leq j\leq n-1)$  
\bea
q^\l&\approx &q^\l(q^0,q^1,...,q^{n-1},p_0,p_1,...,p_{n-1}),
\nn\\&&\hskip 30mm
(n\leq \l \leq 2n-1).
\label{qmm}
\eea
They are combined with the constraints $\vp_\l,\;(n\leq \l \leq 2n-1)$
\bea
\vp_{\l}=p_\l\approx 0,\;\;(n\leq \l \leq 2n-1)
\eea
to form a second class set and can be used to eliminate the canonical pairs
 $\{q^\l,p_\l\}\;(n\leq \l\leq 2n-1)$.

If we take into account \bref{qmdot}
the constraint \bref{EOMm} for $m=0$ is the Euler-Lagrange
equation for the original higher derivative Lagrangian,
\bea  
\psi^0&\sim&\;\sum_{\l=0}^n~ (-\pa_t)^\l
 \frac{\pa L(t)}{\pa(\pa_t^\l q(t))}=0.
\label{EOM0}
\eea
The  constraints \bref{EOMm} for $m>0$ are the time derivatives of the   
Euler-Lagrange equation \bref{EOM0} expressed in terms of $q$'s. 
For a non-singular theory, all the constraints \bref{EOMm} can be rewritten as
\bea
q^\l&-&q^\l(q^0,q^1,...,q^{n-1},p_0,p_1,...,p_{n-1})\approx 0,\;\;\;\;
( \l \geq 2n)
\label{qmmm}
\eea
and can be paired with the constraints $\vp_{\l},\;(\l\geq 2n)$
\bea
\vp_{\l}=p_\l\approx 0,\;\;(\l\geq 2n)
\eea
 forming second class constraints.
They are used to eliminate canonical paires $\{q^\l,p_\l\}\;(\l\geq 2n)$.

In this way the infinite dimensional phase space is reduced to a finite
 dimensional one. 
The reduced phase space is coordinated  by 
$T^*J^n=\{ q^l,p_l\}$ with $l=0,1,...,n-1$.
All the  constraints are second class and 
we use the iterative property of Dirac bracket.
The Dirac bracket for these variables have the standard form,  
\bea
\{q^m,p_n\}^*&=&{\D^m}_n,~~\{q^m,q^n\}^*=\{p_m,p_n\}^*=0.
\label{symplectichh}
\eea
 The Hamiltonian \bref{h} in the reduced space is given by
\bea
H(t)&=&\sum_{m=0}^{n-1}~p_m(t)~q^{m+1}(t)~-~L(q^0,q^1,...,q^n)
\label{hhh}
\eea
where $q^n$ is expressed using \bref{qmm} as a function of the reduced 
variables in $T^*J^n$.
Note that if we consider the limit $n$ going to infinity
the constraints \bref{vpmm} and \bref{EOMm} do not allow, in general, 
to reduced the dimensionality of the infinite dimensional phase space of 
the non-local system via Dirac brackets.

\section{ Symplectic formulation of the Euler-Lagrange equation}

The Hamiltonian formalism presented in the last sections can be
cast into a symplectic form as follows. The Poisson brackets (\ref{PB}) 
correspond to the symplectic two form $\Omega\in\Lambda^2(T^\ast J)$:
\be
\Omega= \int d\lam\,  \D P(t,\lam)\wedge \D Q(t,\lam),
\label{Omega1}
\ee
where $\D$ stands for the functional exterior derivative.

In the constrained phase space  $\Gamma_1 \subset T^\ast J$ defined by
(\ref{mom}) only, the induced (pre)symplectic form is: 
\be
\Omega_1= \frac12 \,\int d \lam\, d\lam^\prime \, \w(t;\lam,\lam^\prime,[Q])\,
 \D Q(t,\lam) \wedge \D Q(t,\lam^\prime) 
\label{Omega}
\ee
where
\be
\w(t;\lam,\lam^\prime,[Q]) = \chi(\lam^\prime,-\lam)\,\int 
d \s \,\frac{\D\CE(t;\s,\lam)}{\D Q(t,\lam')} \,. 
\label{omega}
\ee
The induced Hamiltonian is:
\be
H_1(t,[Q]) = \int d\lam ~F(t,\lam,[Q])\, Q^\prime(t,\lam) - \tilde L(t,[Q]).
\label{HamQ}
\ee
The generator of the time evolution (\ref{timeEv1}) is the vector field
\be
X([Q]) = \int d\lam\, \dot Q(t,\lam)\, \frac{\D \hspace*{2em}}{\D Q(t,\lam)}.
\label{timeGen}
\ee

Now, \hspace*{1em}  $i(X)\W_1+\D H_1=0$ \hspace*{1em}
gives a first order formulation of the EL equation. 
Indeed a short calculation yields
\bea
& & i(X)\W_1+\D H_1~ =~ -\int d \s 
 \CE(t;\s ,0)\,{\D Q(t, 0)} \nonumber \\
& & +\int d\lam d\lam^\prime~\D Q(t,\lam^\prime)
 \left[\dot Q(t,\lam) - Q^\prime(t,\lam)\right] \omega(t;\lam,\lam^\prime).
\label{iXW}
\eea
Whence, the evolution \bref{Qdot} and the EL equation \bref{EL} follow
from it.

\section{ Path integral quantization}

Let us consider the Hamiltonian path integral quantization of the  $1+1$ 
dimensional field theory associated with the \ham (\ref{h}) for
$L^{\rm non}(t)$. The path integral is given by
\bea
&&\int [d P(t,\lam)][dQ(t,\lam)]~\mu
\nn\\ &&~~~~~~~~~~~e^{i\int dt d\lambda (
P(t,\lam)[\dot Q(t,\lam)-Q'(t,\lam)]+\tilde L(t)\delta(\lam))}.
\label{pathh}
\eea
The integration is performed over the reduced phase space 
$\Gamma$ and the measure \onlinecite{faddeev,Senjanovic} $\mu$ is
\bea
\mu&=&\det\pmatrix{\{\vp,\vp\}&\{\vp,\psi\}\cr
                    \{\psi,\vp\}&\{\psi,\psi\}}~\D(\vp)\D(\psi).
\eea
First we consider the non-singular higher derivative
\lag system of order $n$. From the discussions of section IV 
the constraints are arranged to a set in which the canonical variables
$(q^j,p_j)$ for $j\geq n$ are expressed in terms of ones for $0\leq j< n$.
The  measure becomes 
\bea
\mu&=&      \prod_{j=n}^{2n-1}\{\D(p_j)\D(q^j-...)\}
    \prod_{k=2n}^{\infty}\{\D(p_k)\D(q^k-...))\}
\eea
where ... terms are given functions of $(q^i,p_i),(i=0,...,n-1)$.
Integrating over $(q^i,p_i),(i\geq n)$ \bref{pathh} becomes
\bea
&&\int \prod_{i=0}^{n-1}dq^idp_i 
e^{i\int dt \sum_{i=0}^{n-1}p_i(\dot q^i-q^{i+1})
+ L(q^0,...,q^{n})}
\label{pathh2}
\eea
where $q^{n}$ is given as a function of $(q^i,p_i),(i=0,...,n-1)$.
This is the \ham path integral of the Ostrogradski formalism. 
If we assume non-local systems can be regarded as infinite $n$ limit
of higher derivative system \bref{pathh2} becomes, by taking $n\to\infty$, 
\bea
&&\int [d P(t,\lam)][dQ(t,\lam)]~
\nn\\ &&~~~~~~~~~~~e^{i\int dt d\lambda (
P(t,\lam)[\dot Q(t,\lam)-Q'(t,\lam)]+\tilde L(t)\delta(\lam))},
\label{pathh3}
\eea
where $Q$ and $P$, which are  $n\to\infty$ of $(q^i,p_i),(i=0,...,n-1)$,
are not restricted by the constraints in contrast to \bref{pathh}.

If we integrate out the momenta and using $\D(\dot Q(t,\lam)-Q'(t,\lam))$
we get 
 \be
\int [d q(t))] e^{i\int dt ~L^{\rm non}(t)},
\label{pathh1}
\ee
which is the Lagrangian path integral formulation for the non-local theory.

\section{ Application to space-time non-commutative $\phi^3$ theory}

Space-time non-commutative theories have  
peculiar properties due to their acausal behavior and lack of unitarity.
Here we would like to use the previous 
formalism to study the question of unitarity in these theories. 

To fix the ideas we consider a 
non-commutative $\phi^3$ 
theory with arbitrary non-commutativity in $d$ dimensions. The \lag
density is given by  
\bea
\CL^{\rm non}(x^\mu)&=& \frac12\pa_\mu \phi( x )\pa^\mu \phi(x)-
\frac{m^2}{2} \phi(x )^2
\nn\\&&-\frac{g}{3!}
\phi(x )* \phi(x )* \phi(x )
\label{p3lag}\eea
where $*$ is the star product defined by using a general 
anti-symmetric background
$\T^{\mu\nu}$, 
\bea 
f(x)*g(x)=[e^{i\frac{\T^{\mu\nu}}{2}\pa_\mu^\A\pa_\nu^\B} f(x+\A)g(x+\B)]_
{\A=\B=0}.
\eea

The EL equation is 
\bea
(\Box-{m^2})\phi(x)
-\frac{g}{2!}~\phi(x)*\phi(x)~=~0.
\label{ELeq2}
\eea
$x^0$ in \bref{p3lag}-\bref{ELeq2} will be denoted as $t$ hereafter.
We introduce a "new coordinate $x^0$" which plays the role of $\lam$ in the
previous discussion and
introduce the field $Q(t,x^\mu)$ in $d+1$ dimensions. 
Now $t$ is regard  as ``evolution time" and
$x^\mu:=(x^0,\vec x)$ is a 
continuous Lorentzian index. 
Our metric conventions are $\h_{tt}=\h_{00}=-1,\h_{ii}=+1$.
The relation (\ref{timeEv1}) in this case is
\be
Q(t,x^0,\vec x) = \phi(t+x^0, \vec x). 
\label{Qfi}
\ee
The \lag density in $d+1$ dimensions for
$Q(t,x^\mu)$, (see eq.\bref{lagr1}), is
\bea
\CL(t,x^\mu)&=&-\frac12\pa_\mu Q(t,x)\pa^\mu Q(t,x)-
\frac{m^2}{2} Q(t,x)^2
\nn\\&&~-~\frac{g}{3!}
Q(t,x)* Q(t,x)* Q(t,x),
\eea
where now the derivatives in $*$ are with respect to $x^\mu=(x^0,\vec x)$. 
Note that this \lag density is local in the evolution time  $t$.

The momentum constraint \bref{mom} is given by
\bea
&&\vp(t,x^\mu)= 
P(t,x^\mu)-\D(x^0)~Q'(t,x)
\nn\\&&~~~~~+\frac{g}{2!}\int d x'\chi(x^0,-x'^0)
\int d y_1 dy_2 
\nn\\&&K(y_1-x',y_2-x',x-x') Q(t,y_1) Q(t,y_2),
\eea
where $~Q'(t,x)$ denotes $\pa_{x^0}Q(t,x^\mu)$. $K$ is the symmetric
kernel of three
star products,
\bea
&&f(x)*g(x)*h(x)=\int d y_1 dy_2 d y_3 
\nn\\&&~~~~~~K(y_1-x,y_2-x,y_3-x)f(y_1)g(y_2)h(y_3). 
\eea

The \ham \bref{h} is 
\bea
&&H(t)=\int dx~[~P(t,x)~Q'(t,x)~-~\CL(t,x)~\D(x^0)]
\nn\\
&&=\int d x[~P(t,x)Q'(t,x)
\nn\\&&
+\D(x^0)\{-\frac12 Q'(t,x)^2+
\frac12(\nabla Q(t,x))^2+\frac{m^2}{2} Q(t,x)^2
\nn\\&&~~~~~+\frac{g}{3!}Q(t,x)* Q(t,x)* Q(t,x)\}].
\label{h1}
\eea 
The Hamilton equations are
\bea
&&\dot Q(t,x)=Q'(t,x),
\label{eqQ}\\
&&\dot P(t,x)=P'(t,x)-\D'(x^0)[Q'(t,x)]_{x^0=0}
\nn\\&&~~~~~+\D(x^0)\{\nabla^2 Q(t,x)-{m^2}Q(t,x)\}
\nn\\&&~~~~~-\frac{g}{2!}\int dx'
d y_1 dy_2 \D(x'^0) 
\nn\\&&~~~~~K(y_1-x',y_2-x',x-x')Q(t,y_1)~ Q(t,y_2).
\label{eqP}\eea
The stability of the constraint implies the new constraints \bref{constr2} 
\bea
\psi(t,x)&\equiv&
(\nabla^2 -\pa_{x^0}^2-{m^2})Q(t,x)-\frac{g}{2!}~Q(t,x)*Q(t,x)
\nn\\&=&0,~~~at~~~x^0=0.
\eea
By requiring further consistency we have an infinite number of constraints
which can be  written as \bref{constr3}
\bea
\psi(t,x)&=&0~~~~~~{\rm for}~~~-\infty<x^0<\infty.
\label{vp2}
\eea

Using Hamilton equation \bref{eqQ} for $Q$,
\bref{vp2} becomes the EL equation
\bea
(\nabla^2 -\pa_{t}^2-{m^2})Q(t,x)
-\frac{g}{2!}~Q(t,x)*Q(t,x)~=~0,
\label{ELeq1}
\eea
where $\pa_{x^0}$ on $Q$ is replaced by $\pa_t$ both in the
first term and in the $*$ product. It is the original non-local EL 
equation \bref{ELeq2}.

If we write the symplectic form and the \ham in terms of 
$Q(t,x)$, eqs. \bref{HamQ} and \bref{Omega}, we have
\bea
\W&=&
\int d x\, \D(x^0)~\D Q'(t,x)\wedge \D Q(t,x)
\nn\\&-&\frac{g}{4}\int d x\,
\D(Q(t,x)*Q(t,x))\ep(x^0) \wedge \D Q(t,x)
\label{Wphi3}
\eea
and
\bea
&&H=\int dx\frac{\D(x^0)}{2}\{Q'(t,x)^2+
(\nabla Q(t,x))^2 +
{m^2} Q(t,x)^2 \}
\nn\\&&~~~~-
\frac{g}{4} \int d x
(Q(t,x)*Q(t,x))\ep(x^0)Q'(t,x).
\label{Hamp3}
\eea
These expressions can be rewritten in terms of $\phi(x)$ using \bref{eqQ}
{\it i.e.} \bref{Qfi}. 
In particular the interaction Hamiltonian
becomes
\be 
H_i=-
\frac{g}{4} \int d x
(\phi(x)*\phi(x))\ep(x^0)\dot \phi(x)
\label{hint0}.
\ee
Note that the occurrence of time derivatives of any order in the
interaction \ham is not forbidden in non-local theories. This
property is clearly not fulfilled by local theories.

Now we can perform the path integral quantization using \bref{pathh1} to obtain
\be
\int [d\phi(x)]e^{\int dx (\frac12\pa_\mu \phi( x )\pa^\mu \phi(x)-
\frac{m^2}{2} \phi(x )^2-\frac{g}{3!}\phi(x )* \phi(x )* \phi(x ))}.
\ee
From which we read the Lagrangian Feynman rules, \cite{TDLee}.
They coincide with the ones used in \cite{gm}.
Therefore,
it follows from \cite{gm} and \cite{agm} that
non-commutative  $\phi^3$ theory with time-like non-commutativity is not
unitary while non-commutative  $\phi^3$  theory 
with light-like non-commutativity is unitary.
\smallskip

{\em Note Added:} Recently, reference \cite{woodd1} has also
considered the Hamiltonian formalism for non-local theories.
\smallskip


We acknowledge discussions with Luis Alvarez-Gaum\'e, Jos\'e Barb\'on,
Jaume Gomis, Esperanza Lopez, Karl Landsteiner and to
Luis Alvarez-Gaum\'e for a careful reading of the manuscript. 
The work of J.G is partially supported by AEN98-0431, GC 1998SGR (CIRIT), 
the work of J.Ll. is supported by DIGICyT, c. no. PPB96-0384 and by IEC(SCF), 
and K.K. is partially supported by the Grant-in-Aid for Scientific Research, 
No.12640258 (Ministry of Education Japan).


\begin{thebibliography}{99}
\bibitem{fn1}{ From now on when we will refer to non-local theories as the 
        ones with an infinite number of time derivatives. }
\bibitem{LV}{The connection with Ostrogradski formalism has also 
been analized in \cite{lv} in the context of symplectic mechanics.}
\end{thebibliography}
\end{document}